\documentclass[11pt]{article}
\usepackage{proteins}
\usepackage{citesupernumber}
\usepackage{graphics}
\usepackage{overcite}
\usepackage{multirow}
\usepackage{float}
\usepackage{lscape}
\usepackage{doublespace}
\makeatletter
    \newcommand\figcaption{\def\@captype{figure}\caption}
    \newcommand\tabcaption{\def\@captype{table}\caption}
\makeatother

\begin{document}
\bibliographystyle{proteins}
\begin{titlepage}
.
\vspace{1.5 in}

\begin{center}

{\LARGE PIPER: An FFT-Based Protein Docking Program with Pairwise Potentials}
\vspace{.1in}

\large
Dima Kozakov $^{1,3}$ , Ryan Brenke $^{2,3}$, Stephen R. Comeau $^{2}$, and Sandor Vajda$^{1,2,*}$\\
\end{center}
\begin{flushleft}
$^{1}$ {\it Department of Biomedical Engineering, 
Boston University, Boston, Massachusetts}\\
$^{2}$ {\it Program in Bioinformatics, 
Boston University, Boston, Massachusetts}\\
$^{3}$ {\it Joint first authors}\\

\end{flushleft}

\begin{flushleft}
{\bf Running title:} Docking with Pairwise Potentials \\
\end{flushleft}

\begin{flushleft}
$^{*}$ {\bf Corresponding author:} Sandor Vajda, Department of Biomedical Engineering \\
Boston University, 44 Cummington street, Boston, MA 02215, USA \\
e-mail: vajda@bu.edu,  phone: 617-353-4757,  fax : 617-353-6766 \\
\end{flushleft}

\end{titlepage}
{\bf{ABSTRACT}} 
The Fast Fourier Transform (FFT) correlation approach to protein-protein docking can evaluate the energies of billions of 
docked conformations on a grid if the energy is described in the form of a correlation function. Here, this restriction 
is removed, and the approach is efficiently used with pairwise 
interactions potentials that substantially improve the docking results. The basic idea is approximating the 
interaction matrix by its eigenvectors corresponding to the few dominant eigenvalues, resulting in an energy 
expression written as the sum of a few correlation functions, and solving the problem by repeated FFT calculations.
In addition to describing how the method is implemented, we 
present a novel class of structure based pairwise intermolecular potentials. The DARS (Decoys As the Reference State) 
potentials are extracted from structures of protein-protein complexes and use large sets of docked conformations as 
decoys to derive atom pair distributions in the reference state. 
The current version of the DARS potential works well for 
enzyme-inhibitor complexes. With the new FFT-based program, DARS provides much better docking results than the earlier 
approaches, in many cases generating 50\% more near-native docked conformations. Although the potential is far 
from optimal for antibody-antigen pairs, the results are still slightly better than those given by an earlier FFT 
method. The docking program PIPER is freely available 
for non-commercial applications. 


\newpage
{\bf INTRODUCTION} \\
The goal of protein-protein docking is to determine the structure of a complex in atomic detail, 
starting from the coordinates of the unbound component molecules \cite{nussinov, smith, vajdarev}.  
Most of the current docking methods start with rigid body docking that generates a large number of 
docked conformations with good surface complementarity \cite{glass}. The Fast Fourier Transform (FFT) 
correlation approach, introduced in 1992 by Katchalski-Katzir and co-workers \cite{Katchalski}, 
revolutionized this step of rigid body search. Owing to the numerical efficiency of this algorithm it 
became computationally feasible, for the first time, to systematically explore the conformational space of protein-protein 
complexes evaluating the energies for billions of conformations on a grid, and thus to 
dock proteins without any a priori information on the expected structure \cite{wodak1, wodak2}. 
Other approaches, primarily Monte Carlo, also perform well if the search can be restricted to regions of 
the conformational space 
\cite{ruben, Baker}, but become computationally expensive if no such constraints are available. 
For this reason, FFT-based docking is  the first step in many methods that have performed well 
at CAPRI (Critical Assessment of Predicted Interactions), the first community-wide experiment 
devoted to protein docking \cite{wodak1, wodak2}. We note that this approach is obviously restricted to 
proteins with moderate conformational changes upon binding \cite{glass}.

While the FFT based method represents major progress in protein docking, 
it also has serious limitations, even beyond the consequences of the rigid body assumption. 
The most important constraint is on the target function, which is restricted to have the form of a 
correlation function, resulting in rather inaccurate estimation of the binding free energy. 
The original scoring function, introduced by Katchalski-Katzir et al. \cite{Katchalski}, was based 
only on shape complementarity, but was later extended to include additional terms representing 
electrostatic interactions \cite{Gabb,mandell}, or both electrostatic and solvation contributions 
\cite{ZDOCK}. While the new potentials improved performance, energy evaluation remains relatively 
crude. Due to this uncertainty, to avoid the loss of near-native solutions when docking unbound structures of proteins, 
one has to retain a large number 
(usually 2,000 to 20,000) of docked conformations for further analysis. Since the number of near-native structures 
among the ones retained is generally small -- from a few to at most a hundred -- the rigid body search ends with 
many false positives \cite{glass}, i.e., conformations that are geometrically distant from the native but score as well as 
the ones close to it.
Accordingly, in the best performing docking methods the initial search is followed by a refinement and discrimination 
step that ranks the docked 
conformations and selects the ones close to the native, usually using a more accurate energy function that accounts for 
the affinity of binding between the two proteins \cite{Gabb, Gatchell, RDOCK}. The discrimination between 
near-native and other structures can be further improved by clustering methods \cite{ClusPro,cluster}. 
These procedures improve the discrimination such that conformations with less
than 10 \AA\  RMSD are 
generally found within the top ten to hundred structures. However, it is needless to say that the discrimination 
step is difficult if the rigid body search generates only a few 
near-native structures, and it is obviously futile if no such structures are in the set. 
Thus, improving FFT methods remains the key to the success of the entire procedure that 
starts with rigid body docking.

In this paper we explore the use of pairwise structure-based potentials with FFT correlation 
docking. Such potentials (also called knowledge-based or statistical potentials) have emerged as  
powerful tools for finding near-native conformations in sets of structures generated by search 
algorithms in macromolecular modeling, and have substantially contributed to improving the accuracy 
in protein structure prediction \cite{Skolnick_fold, Lu_fold, Godzik, Jernigan1, Jernigan2, 
Subramaniam, DFIRE}. Pairwise knowledge-based potentials have also been used with success in the discrimination 
stage of protein-protein docking \cite{Gatchell, RDOCK, ClusPro, RPScore, Murphy, Lu_dock, Zhou_dock}. 
Hence their use directly in the docking stage is expected to increase the number of near-native 
structures found. In  principle, FFT based methods can use pairwise potentials as their scoring function.
A potential defined for $K$ atom types and given by a $K \times K$ interaction matrix can be written 
as the sum of $K$ correlation functions. The function can be then evaluated by performing $K$ 
forward and $K$ inverse Fourier transformations. The major difficulty is that $K$ is generally around 20, 
and hence the approach 
is computationally expensive, even with the increasing 
computer power currently available. Here we show that this 
problem can be avoided by an eigenvalue-eigenvector decomposition of the coefficient matrix that 
substantially reduces the complexity of the calculations. In fact, adequate accuracy can be achieved 
by restricting consideration to the eigenvectors corresponding to the $P$ largest eigenvalues where 
$ 2 \leq P \leq 4$, 
and thus performing only 2 to 4 forward and the same number of inverse FFT calculations. According to the results presented in the paper, 
this approach substantially increases the number of near-native solutions (hits) at relatively 
moderate additional computational costs.

Although our focus is on the extension of the FFT docking, we also describe preliminary work on developing 
a new class of structure-based potentials. While a large variety of 
intra-molecular potentials are available for protein folding and fold recognition, relatively 
little attention was given to inter-molecular potentials, partly because the number of known 
protein-protein complexes only recently started to grow. Sternberg and co-workers developed both 
residue-level and atom-level inter-molecular pair potentials; a residue-level potential based on $C_{\alpha}$
atoms, a residue-level potential based on all atoms (RPScore), a residue-level potential based on all 
side-chain atoms, and an atom-level potential with $K=40$ grouped atom types \cite{RPScore}. The potentials 
were derived from a small training set that included a few hetero-dimers, a few homo-dimers, and a nonredundant 
set of protein domains. Moont et al. \cite{RPScore} tested the potentials using decoy sets of docked 
conformations only for 9 complexes.  The best discrimination was obtained by the residue-level potentials.  
However, we have observed \cite{Murphy} that RPScore was much more likely to fail for a complex 
that was not represented (directly or by homology) in its training set, suggesting that the dataset was 
too small and biased toward certain types of complexes. Skolnick and co-workers \cite{Lu_dock} also 
developed both a residue-level ($K=20$) contact potential and an atom-level potentials with all $K=167$ 
heavy atoms as atom types. The potentials have been derived from 768 protein complexes (617 homo-dimers 
and 151 hetero-dimers), and were tested using decoy sets of docked conformations for 15 complexes. 
In contrast to the results by Moont et al. \cite{RPScore} , the discrimination turned out to be much better 
using the atom-level potential than using the residue-level version, suggesting that the training set used 
by Moont et al. was simply too small. This is in good agreement with our preliminary results, and hence 
we restrict consideration to atom-level potentials. 

Two atom-level potentials will be used in this work, both written in the form 
$E_{pair} = \sum_{ij} \varepsilon_{ij}$, where $\varepsilon_{ij}$ denotes the energy contribution by a pair
of interacting atoms $a_i$ and $a_j$, and the sum is taken over all pairs of atoms that are closer to 
each other than a
cutoff distance $D$.  The first potential $E_{ACP}$, termed ACP (Atomic Contact Potential) 
\cite{ACE,ACE2}, is an atom-level extension of the well known residue-level potential by 
Miyazawa and Jernigan \cite{Jernigan1, Jernigan2}. ACP is a solvent-mediated potential (see Methods 
and \cite{mediated} for the discussion of this concept), and the atomic contact energy, $\varepsilon_{ij}$,   
is defined as the effective free energy of a reaction in which two fully solvated atoms desolvate and associate 
to form the interacting atom pair $a_ia_j$ \cite{Jernigan1, Jernigan2}.
The ACP potential has been derived for $K=18$ atom types from 89 non-homologous proteins \cite{ACE}. Due to the 
solvent-mediated character and the small number of charged residues in the interior of proteins, the atomic 
contact potentials may be attractive even between atoms with like charges. More generally, since ACP
includes both desolvation and atom-atom binding, the atomic
contact energies between polar or charged atoms are weak. For this reason, we have used the potential 
in conjunction with a coulombic electrostatic term. For docking applications we also added a 
van der Waals term, representing shape complementarity of the component proteins. 

The second atom-level contact potential has been specifically developed for application 
with the FFT docking. As will be described in the Methods, to derive the potential we 
considered the nonredundant data 
set of 621 protein-protein interfaces, compiled by Glaser et al. \cite{ilyaset}, but removed
all complexes that also belong to the benchmark sets \cite{benchmark1,benchmark2} used for testing our method.
Unlike the solvent-mediated ACP, this 
potential is residue-mediated (see Methods and \cite{mediated}).  While in a solvent-mediated potential 
the reference state is defined by non-interacting and fully solvated atoms, in a residue-mediated 
potential the reference state is obtained by averaging interactions over compact structures, most 
frequently by using the mole fractions of specific atom types \cite{mediated, Lu_fold, Godzik}. The 
novelty of our approach is that we generate a large decoy set of docked conformations 
to be used as a reference state. Developing the potential, we  compare the frequency
of contacts between two specific atom types in the native state to the frequency  of contacts in the 
decoys. Since the goal is finding complex conformations close to the native among the many structures that 
all have good shape complementarity, this scoring scheme is 
natural, as it rewards the occurrence in the interface of the atom pairs that are frequently seen to 
interact in native complexes.  

In the Methods section we describe implementing FFT based docking with pairwise potentials, and briefly
the development of the novel DARS (Decoys as the Reference State) potential.  As with ACP, we add 
electrostatic and van der Waals terms to the DARS potential when used for docking. The properties of the newly developed 
DARS potential significantly differ from those of the Atomic Contact Potential (ACP) \cite{ACE}, and the best 
results are obtained when using a linear combination of ACP and DARS as the scoring function. 
For enzyme-inhibitor 
complexes the results are much better than those obtained by a traditional 
FFT method. In particular, the number of near-native docked structures increased by at 
least 50\% for more than half of the enzyme-inhibitor complexes in the well known protein docking 
benchmark sets \cite{benchmark1, benchmark2}. For the antigen-antibody 
test set the results are weaker, but still better than those by an earlier method. 
The difference is not completely surprising, as analyses 
of protein complexes \cite{jones, LoConte, jackson, chakrabarti} clearly show that the interfaces 
in enzyme-inhibitor and antigen-antibody complexes substantially differ. Our 
results further emphasize that to substantially improve the docking of antibodies to antigens 
one needs a special potential accounting for the properties of the interface in this type of complexes. 

{\bf METHODS}

{\bf FFT Docking with Multiple Correlations}\\
Fast Fourier Transform (FFT) docking algorithms perform exhaustive evaluation of simplified energy functions
in discretized  6D space of mutual orientations of the protein partners.
The larger docking partner is considered the receptor and its center of mass is fixed  at
the origin of the coordinate system. The other partner is  
considered the ligand and all its possible orientational and translational
positions are evaluated at the given level of discretization.
The rotational space is sampled using a deterministic layered Sukharev grid sequence
for the rotational group $SO(3)$, which  quasi-uniformly covers the space with a
given number of samples \cite{lavalle}. The translational space is represented as a grid of displacements of
the ligand center of mass with respect to the receptor's center of mass.

Here we assume that the energy-like scoring function describing the receptor-ligand 
interactions is  defined on a grid and is expressed as the sum
of P correlation functions for all possible translations $\alpha,
\beta, \gamma$ of the ligand relative to the receptor
$$
E(\alpha,\beta,\gamma)=\sum_{p}\sum_{i,j,k}{R_{p}(i,j,k)L_{p}(i+\alpha,j+\beta,k+\gamma)}
$$ 
where $R_{p}(i,j,k)$ and $L_{p}(i,j,k)$ are the components of the correlation function defined on the
receptor and the ligand, respectively. This expression can be efficiently calculated using $P$ forward and 
$P$ inverse Fast Fourier transforms, denoted by $FT$ and $IFT$, respectively:
\begin{eqnarray*}
E(\alpha,\beta,\gamma)& =
&\sum_{p}^P IFT \{ FT^{*}(\{ R_{p} \} FT \{ L_p \} \} (\alpha,\beta,\gamma)\\
FT \{ F \} (l,m,n)&=&\sum_{i,j,k} {F(i,j,k)exp^{-2\pi{\textbf{i}} (li/N_{1}+mj/N_{2}+nk/N_{3})}}\\
IFT \{ f \} (i,j,k) &=&{1\over{N_{1}N_{2}N_{3}}}\sum_{l,m,n} {f(l,m,n)exp^{2\pi {\textbf{i}} (li/N_{1}+mj/N_{2}+nk/N_{3})}}
\end{eqnarray*} 
where ${\textbf{i}} = \sqrt{-1}$, $N_{1}$, $N_{2}$, and $N_{3}$
are the dimensions of the grid along the three coordinates. If $N_{1} = N_{2}= N_{3} = N$,
the efficiency of this approach is $O(N^3\log(N^3))$ as compared to $O(N^6)$
when all evaluations are performed directly.
For each rotational orientation,  which is taken consecutively from the
set of rotations, the ligand is  rotated and the $L_p$ function is calculated on the grid.
We then calculate the correlation function of $L_p$ with the pre-calculated $R_{p}$ function using FFT. 
The resulting sum provides scoring function values for all possible translations
of the ligand. The results are clustered with a 10 \AA\ cube size and one or several lowest energy translations for the given 
rotation are reported. Finally, results from different rotations are collected
and sorted.

{\bf Scoring Function}\\
The energy function is given as the sum of terms representing shape complementarity, electrostatic, and 
desolvation contributions, the latter described by a pairwise potential as follows. 
\begin{eqnarray*}
E & = & E_{shape}+ w_{2} E_{elec}+ w_{3} E_{pair}\\
E_{shape} & = & E_{attr}+w_{1}E_{rep}\\
E_{elec} & = & \sum_{i=1}^{N_r}\sum_{j=1}^{N_l}{q_{i}q_{j}\over{\left(r_{ij}^2+D^{2}exp\left({-r_{ij}^{2}\over{4D^{2}}}\right)\right)^{1\over2}}}\\
E_{pair} & = &\sum_{i=1}^{N_R}\sum_{j=1}^{N_L}{\varepsilon_{ij}}
\end{eqnarray*}
where $N_R$ and $N_L$ denote the numbers of atoms in the receptor and the ligand, respectively.
According to these expressions, the shape complementarity term $E_{shape}$
accounts for both attractive and repulsive interactions, the latter eliminating atomic overlaps. 
The specific form of $E_{shape}$ will be defined on a grid in the next section. The 
electrostatic term, $E_{elec}$, is given by a simplified generalized Born type expression. The coefficients 
$w_{1}$, $w_{2}$, and $w_{3}$ weight the different contributions to the
scoring function. The value of $w_{1}$ is selected to avoid substantial steric clashes, but to
allow for some atomic overlaps that occur due the differences between bound and unbound 
(i.e., separately crystallized) structures of the component proteins. We note that all rigid body 
docking methods assume that such differences exist but are moderate. While this assumption is frequently 
acceptable, it excludes the application of the method to certain types of complexes. 
For example, the benchmark sets \cite{benchmark1,benchmark2} include 
a number of ''difficult" cases with substantial backbone conformational changes upon association.
Most docking methods, including ours, provide few if any near-native conformations for these complexes. 
Although we include some ''difficult" cases in our test set, backbone flexibility is beyond the scope of 
this paper, and the problem
will not be further discussed here. The coefficients $w_{2}$ and $w_{3}$ will be selected 
to provide maximum performance for specific classes of 
proteins.

{\bf Shape Complementarity Energy Terms Defined on a Grid} \\
For efficient evaluation we are using a rectangularly smoothed shape complementarity  term as 
suggested by Vakser \cite{vakser1994}. The repulsive interactions are cut off at 
the van der Waals radius $r_{vdw}$ plus 2 \AA\ because we want the penalty function to be 
tolerant enough and to allow for differences between bound and unbound structures. 
To further account for the potential flexibility of the component proteins we have reduced the 
van der Waals radii of atoms 
on the protein surface, and increased the radii in the core. 
The attractive part has the same cutoff radius (6 \AA)  for all atom types. On the grid, the functions 
describing the receptor and the
ligand can be represented as follows
\begin{eqnarray*}
R_{p}(l,m,n)&=&-c_{l,m,n}+w_{1}r_{l,m,n}\\
L_{p}(l,m,n)&=&\left\{\begin{array}{ll} 1 & \hbox{if} \quad (l,m,n) \owns (a_j \in     
    J) \\0 & \hbox{otherwise} 
\end{array}\right.
\end{eqnarray*}
where $(l,m,n) \owns (a_j \in J)$ means that the grid point $(l,m,n)$ overlaps with atom $a_j$ of atom type $J$,
$c_{l,m,n}$ is number of atoms that are at the distance $d<r<D$ from the grid point $(l,m,n)$, and $r_{l,m,n}$
is number of atoms that are at the distance $r<d$ from the same grid point. We have used the values  $D=6$ \AA\
and $d=r_{vdw}+2$ \AA. The correlation of these two functions provides a shape 
complementarity term representing both repulsive and attractive interactions, the former 
for the distances $r<d$, and the latter in the range $d<r<D$. 

{\bf Electrostatic Interactions on a Grid}\\
To account for the electrostatic interactions between the two proteins surrounded by solvent we
use a simplified Generalized Born (GB) type equation with constant Born
radii. This approximation neglects the dependence of the Born radii on the atomic environment, but allows for writing
the electrostatic interactions as a correlation between the electrostatic potential field of the receptor and 
the charges on the ligand:
\begin{eqnarray*}
R_{p}(l,m,n) & = &
\sum_{i=1}^{N_r}{q_{i}\over{\left(\hat r_{i,(l,m,n)}^2+D^{2}exp\left({- \hat r_{i,(l,m,n)}^{2}\over{4D^{2}}}\right)\right)^{1\over2}}}\\
L_{p}(l,m,n)&=&\left\{\begin{array}{ll} q_{j}& \hbox{if} \quad (l,m,n) \owns (a_j \in     
    J) \\0 &
    \hbox{otherwise}\end{array}\right.\\
\hat r_{i,(l,m,n)}&=&max(r_{i,(l,m,n)},2D)
\end{eqnarray*}
where $r_{i,(l,m,n)}$ is the distance between atom $a_i$ and the grid point $(l,m,n)$. The potential is truncated 
at the distance $2D$ for the same reason as the shape complementarity term. 
In addition, the electrostatic interactions are made less sensitive to conformational perturbations by 
smoothing it through a convolution with square boxes of size 3 \AA\ . This type of smoothing is very important. 
As shown in Figure 1A, the function $R_{p}(l,m,n)$ yields a very rugged electrostatic potential field where 
the positions of the local minima and maxima heavily depend on the atomic coordinates. The convolution with the box 
yields a much smoother potential (Fig. 1B), which is less sensitive to coordinate perturbations. 
The same applies to the electrostatic part of the receptor-ligand interaction energy. Figures 1C shows a slice of 
this energy, calculated with the original electrostatic potential, as a function of two translational coordinates. 
Figure 1D shows the same slice of the energy, but this time calculated using the smoothed potential. 

{\bf Corellation Decomposition of Pairwise Potentials}\\
In general form  of a pairwise contact potential is
$$
E_{pair}=\sum_{i=1}^{N_R}\sum_{j=1}^{N_L}{\varepsilon_{ij}}
$$
In this equation $N_R$ and $N_L$ denote the numbers of atoms in the receptor and the ligand, 
respectively; $\varepsilon_{ij}= \varepsilon_{IJ}$ if the interacting atoms 
$a_i$ and $a_j$ are of types $I$ and $J$, respectively, and $d< r_{ij}<D$; whereas $\varepsilon_{ij}=0$ if $r_{ij}>D$.
Here $\varepsilon_{IJ}$ is the contact energy between interacting atoms of types $I$ and $J$. 
The above expression for $E_{pair}$ does not have the form of a correlation function, but it can
be written as a sum of correlation functions. This latter representation is based on the 
eigenvalue - eigenvector decomposition of the pairwise interaction matrix of the elements $\varepsilon_{IJ}$.
The matrix is symmetric and hence 
has $K$ real eigenvalues, where $K$ is the number of different atom types. The matrix elements can be 
written as 
$$
\varepsilon_{IJ} = \sum_{p=1}^{K}{\lambda_{p}u_{pI}u_{pJ}}
$$
where $\lambda_{p}$ is $p$th eigenvalue of the interaction matrix, and $u_{pI}$ is the $I$th component 
of the $p$th eigenvector. Thus, any pairwise potential can be
calculated using $K$ real or $K/2$ complex FFTs. Since the existing pairwise interaction potentials 
can have up to 167 atom types, the calculation can be computationally very
expensive. However, we can approximate the total pairwise energy with arbitrary accuracy using a 
much simpler expression. Each term 
in the eigenvalue - eigenvector decomposition represents an energy contribution 
proportional to the absolute value of the 
eigenvalue $\lambda_{p}$, and such contributions are independent due to the orthogonality of the eigenvectors. 
We order the eigenvalues by their absolute values, starting with the largest, and restrict 
consideration to the first $P$ terms, i.e., neglect the contribution of the remaining terms. 
Note that restricting consideration to a grid may yield up to 10 \% error in the energy values,  
and hence it is well justified to truncate the summation when the energy
contributions of the neglected terms are comparable to this error. We have performed
analysis on several existing pairwise potentials such as ACP \cite{ACE} and RPScore \cite{RPScore}, 
and found that, depending on the number of atom types, only 
2 to 4 eigenvalues are needed to achieve this accuracy (see Results). The energy 
term with the $p$th eigenvalue of the pairwise potential on the grid is represented by the function 
\begin{eqnarray*}
R_{p}(l,m,n)&=&\sum_{i=1}^{N_r}{u_{pI}\delta_{i}}\\
L_{p}(l,m,n)&=&\left\{\begin{array}{ll}u_{pJ}& \hbox{if} \quad (l,m,n) \owns (a_j \in     
    J) \\0 & \hbox{otherwise} \end{array}\right.
\end{eqnarray*}
 where $\delta_{i}$ is 1 if the grid point $(l,m,n)$ is at a distance less then D from atom i
 of the receptor. 
 
{\bf Parameters for Enzyme-Inhibitor and Antigen-Antibody Complexes}\\
Most parameters of the FFT algorithm are independent of the type of the  proteins to be docked. 
We sampled 70,000 rotations which aproximately corresponds to sampling at 
every 5 degrees in the space of Euler angles. Increasing the
number of rotations generally improved the results. Thus, the number of points 
was chosen as a compromise between performance and computational efficiency. We used grids with  
1.2 \AA\ cell size, which was found to be 
adequate  for representing protein structures with sufficient 
details and at the same time providing acceptable computational efficiency. 
The number of grid cells along each direction was selected on the basis of the size of the
receptor and the ligand by the following algorithm: 
\begin{eqnarray*}
a(v)&=&abs(v_{i}-v_{j})\\
S_x&=&max_{ij}(a(x_{r}))+max_{ij}(a(x_{l}),a(y_{l}),a(z_{l}))\\
S_y&=&max_{ij}(a(y_{r}))+max_{ij}(a(x_{l}),a(y_{l}),a(z_{l}))\\
S_z&=&max_{ij}(a(z_{r}))+max_{ij}(a(x_{l}),a(y_{l}),a(z_{l}))\\
N_x&=&r((S_x/1.2)+2)\\
N_y&=&r((S_y/1.2)+2)\\
N_z&=&r((S_z/1.2)+2)
\end{eqnarray*}
where $N_x, N_y,$ and $N_z$ denote the number of cells along the x, y, and z directions, respectively, 
and $r(N)$ is a function which moves $N$ to the nearest product of small prime
numbers. The algorithm selects the smallest grid that can accommodate the two 
proteins, and is efficient for the Fast Fourier Transform. It is interesting to  
note that the grid size we have used was actually too small to fit the whole ligand if
the ligand center of mass was shifted to the far ends of the grid. Since the grid is
assumed periodical in the FFT calculations, such ligands are effectively wrapped around the receptor. 
However, this effect occurs only at substantial separations of the two proteins where the interactions  
are weak, and hence do not effect the calculated energy values.\par

As will be discussed, the ACP potential works well for complexes with a largely hydrophobic interface, but 
the DARS potential provides better discrimination if the interface is more polar. Since the properties of the 
interface are not a priori known, we use the linear combination of the two potentials defined by 
$$
E_{DARS+ACP}=(3*E_{DARS}+0.5*E_{ACP})/4.0
$$
as the pairwise potential $E_{pair}$ in the FFT-based docking calculation. 

The parameters that differ between different types of complexes are the weights $w_1$, 
$w_2$, and $w_3$ of the energy terms in the energy expression. These parameters were optimized 
and adjusted using a small subset of benchmark proteins \cite{benchmark1} taken from 
the Protein Data Bank (PDB). For enzyme-inhibitor pairs
we have used 
the complexes 1ACB, 1BRC, 1DFJ, 2KAI, and 4HTC, whereas the complexes
1WEJ, 1AHW, 1E08, and 1NCA were used to find appropriate weights for docking antibody and antigen pairs.
For each complex, 20,000 docked conformation were generated 
using the FFT algorithm with some initial values of the weights. The resulting structures were divided 
into two subsets, one with conformations within 10 \AA\ RMSD from the native structure, and the other beyond 
this RMSD cutoff.  We then used logistic regression as provided in 
the package R (see http://www.r-project.org/), and optimized the weighting coefficients. 
This was done several times iteratively to achieve convergence. Based on these calculations
we used the values $w_1=2.1, w_2=133.0$, and $w_3=2.2$ for enzyme-inhibitor complexes, and
$w_1=2.0, w_2=400.0$, and $w_3=1.0$ for antigen-antibody pairs. Thus, the coefficient $w_1$ of the 
repulsive contribution in the shape complementarity term turned out to be essentially independent of the 
type of the complex. We have found, however, that the optimal weight $w_2$ of the 
electrostatic component is three times larger in antigen-antibody than in enzyme-inhibitor complexes, 
in agreement with the fact that the latter complexes generally have a less polar interface. 

As will be described in the Results, test calculations were performed on 33 enzyme-inhibitor 
complexes found in the protein docking 
benchmark sets 1.0 and 2.0 \cite{benchmark1,benchmark2}, and for 16 antigen-antibody pairs 
found in the benchmark set 1.0. For each complex we docked the unbound-unbound or the unbound-bound structures 
of the component proteins as available in the benchmark sets \cite{benchmark1,benchmark2}.
In both the receptor and the ligand, we masked the attractive shape complementarity terms
for the terminal residues. The reason is that the position of these 
residues is frequently uncertain, which may lead to false positive interactions. 
For antibodies we also masked the attractive shape term for all residues that did not 
belong to the Complementarity 
Determinig Regions (CDRs), see Chen et al. \cite{zdock1}.

{\bf Development of Potentials with Decoys as the Reference State (DARS)}\\
Within the framework of the inverse Boltzmann approach, a statistical 
potential between two atoms $a_i$ and $a_j$ that are of types $I$ and $J$, respectively, and are located within a 
certain cutoff distance $D$, is defined by the expression of the form 
$$
\varepsilon_{IJ} = -RT\ln(p_{IJ})
$$
where R is the gas constant, T is the temperature, and $p_{IJ}$ denotes the probability 
of two atoms of types $I$ and $J$ interacting. 
This  probability is approximated by the frequency
$$
p_{IJ}={{{\nu^{obs}}_{IJ}}\over{{\nu^{ref}}_{IJ}}}
$$
where ${\nu^{obs}}_{IJ}$ is the observed number of interacting atom pairs if types $I$ and $J$,  
and ${\nu^{ref}}_{IJ}$ is the expected number of interacting atom pairs of types $I$ and $J$ assuming an 
appropriate reference state. If the state of the protein or complex is fully determined 
by the interactions among its interactions sites, the (structure-based, knowledge-based, 
or statistical) potential of the system is calculated by the sum 
$E_{pair}  = \sum_{i} \sum_{j} \varepsilon_{ij}$, where 
$\varepsilon_{ij}= \varepsilon_{IJ}$ if the interacting atoms 
$a_i$ and $a_j$ are of types $I$ and $J$, respectively, and $d< r_{ij}<D$; 
whereas $\varepsilon_{ij}=0$ if $r_{ij}>D$. 

The basic idea of knowledge-based potential is that ${{\nu^{obs}}_{IJ}}$ can be directly 
determined by 
counting the number of intermolecular interactions between atoms of types $I$ and $J$
in a database of protein complexes. 
The advantages of structure-based potentials are clear.  The potentials include the essential 
features of intermolecular interactions as well as solvent effects. 
Because the potentials are fast to compute, they allow better sampling of the conformational 
space in the calculations. Since the numbers of known protein complex structures have increased 
greatly in recent years, and certainly will grow even faster in the near future, these potentials 
are expected to become more and more accurate if the additional structural information is properly utilized.
The selection of the reference state remains a critical feature. The general assumption 
in the reference state is that the specific interactions determining the distribution 
of interaction sites are removed as much as possible. Since experiments do not provide 
us with such ``random'' protein complexes, additional assumptions have to be made, and this is the 
point where the various structure-based potentials start to differ 
\cite{Godzik}. On the basis of the reference state, 
structure-based potentials can be divided into two large groups.
\cite{mediated, Godzik, Jernigan_new}
  
In solvent-mediated potentials the reference state is defined in terms of solvated but 
otherwise non-interacting residues \cite{mediated}. The advantage of this approach 
is that the reference state has some physical meaning, i.e., for an intermolecular 
potential the reference state is defined by solvated component proteins at infinite 
separation. By definition, solvent-mediated potentials are required for estimating the 
binding free energy, and thus evaluating the strength of the association. 
Due to the finite size of proteins, inter-residue distances in complexes are relatively 
short even for residue pairs that might repel each other. Since these effects are not 
compensated by the reference state, solvent-mediated potentials may be attractive 
even for two interacting residues with charges of the same sign \cite{Jernigan1}. 
By definition, the Atomic Contact Potential (ACP), one of the target functions we use in our 
FFT calculations, is a solvent-mediated 
potential \cite{ACE}. Because of this and because it has been derived 
from protein structures in which salt bridges are rare, ACP essentially fails to represent 
the electrostatic interactions. Nevertheless, it performs well for complexes 
in which the interface is largely hydrophobic, which is the case in the majority of 
enzyme-inhibitor complexes. 

In residue-mediated potentials the reference state is obtained by averaging 
the interactions over compact structures \cite{furuichi, mediated}. Since we observe the unfavorable pair interactions 
less frequently than in the reference state, the corresponding contributions 
to the potential are repulsive as they should be. Such potentials are more 
suitable for finding near-native conformations in a set of compact structures 
than the ones based on the solvent-mediated approach. The disadvantages are 
that averaging generally involves an ensemble of compact conformations that are non-physical, 
and the derivation requires additional assumptions. The most frequently used reference 
state uses the mole fractions to define $ {\nu^{ref}}_{IJ} = \nu^{obs}\times X_I \times X_J$,
where $\nu^{obs}$ is the total number of interacting pairs with the distance constraints
$d < r_{ij} < R$, and $X_I$ is the mole fraction of atom type $I$, 
defined as $\nu_I/\nu$, i.e., the atom composition of the entire complex 
was used to normalize the number of expected interactions.  
A similar approach is based on the same formulas, but considering only the atoms in some 
neighborhood of the interface when calculating the reference frequencies. Both approaches 
assume that the reference state contains a random mixture of atoms in volumes that correspond 
either to the complex or to the interface region. 

Most residue mediated potentials have been derived from folded  protein 
structures, and were primarily used for finding near-native conformations among protein structures generated 
by some prediction algorithm, involving searches in a large conformational space 
\cite{Skolnick_fold,Lu_fold,Jernigan1,Subramaniam}.
Since rigid body protein-protein docking requires searching only in six dimensions, 
it is feasible to generate large sets of docked conformations. Using only the van der Waals interaction term 
as the target function, the resulting conformations 
do not depend on specific atomic interactions and hence 
are essentially random complexes, but with good shape complementarity.
Thus, frequencies of atom pair interactions in the reference state 
can be obtained by counting the specific atom-pair interactions in such decoy sets. Since our goal 
is finding 
docked structures with high levels of ''chemical" complementarity among the many compact structures 
generated by the FFT algorithm, it is natural to define the probability for an atom pair as the
frequency of the pair in the native complexes, divided by the 
frequency of the same pair in the decoys. The frequencies are normalized using the total numbers of 
atom pairs in the native structures and in the decoys, respectively. Hence the relative sizes 
of native and decoy sets do not matter, provided that they are large enough to yield appropriate 
statistics. We note that an approach somewhat similar to DARS has been developed by Bernauer et al. 
\cite{janin}, who considered the native states of about 80 protein complexes, together with about 100 
decoys for each complex.  However, rather than the decoys providing the reference distribution 
function as in the current work, an evolutionary learning program was used to generate a scoring 
function to separate the natives and the decoys. While the approach needs further refinement, it 
shows that it is possible to generate meaningful decoy sets of docked protein structures, and to use 
the properties of these decoys for discrimination.

{\bf RESULTS AND DISCUSSION} 

{\bf Developing and Testing DARS Potentials}\\
Developing the structure based DARS potential requires the selection of atom types, 
the definition of interactions (i.e., distance cutoff values), a training set of native complexes, 
and a decoy set of docked complexes for the reference state. 
Since we also used the Atomic Contact Potential (ACP) in the FFT calculations, 
for simplicity we 
adopted the same 18 atom types defined by Zhang et al. in ACP \cite{ACE}. While the classification of 
atoms was somewhat intuitive, it was generally based on considerations 
of chemical properties and interactions. A detailed description 
of the 18 atom types is given in the original ACP paper \cite{ACE}.
Here we note only that the backbone atoms are considered as separate atom types N, CA, C, and O, 
whereas most hydrophobic side chain atoms are in the $FC^{\zeta}$ and $LC^{\delta}$ categories 
(Table I). We have used the cutoff 
distance of 6.5 \AA\ when counting the frequencies of atom-atom interactions. 

As the training set, we have used a nonredundant database of native protein-protein 
complexes collected by Glaser et al. \cite{ilyaset} from the Protein Data Bank (PDB). The original set 
included 621 protein interfaces from 492 PDB entries. The nonredundant character of this 
database was assured by excluding proteins with more than 30\% sequence identity. While nonredundant, the 
database is far from unbiased in terms of representing protein-protein complexes. In fact, of the 
621 interfaces, 404 are from homodimers. In addition, the set includes a number of enzyme-inhibitor 
and antibody-antigen complexes, 
and few other types. As will be described, we use the protein-protein benchmark sets by \
Chen et al. \cite{benchmark1} and by  Mintseris et al. \cite{benchmark2} for testing the docking algorithm, 
and hence the complexes in 
these benchmark sets were removed from the training set, resulting in 583 interfaces from 
466 protein entries. This set clearly overrepresents oligomeric proteins. The 
fraction of enzymes-inhibitor complexes is also high. Thus, the resulting potential is 
expected to work best for oligomeric proteins and enzyme-inhibitor complexes. Since structures of the separate
subunits in oligomeric proteins are rarely determined, in this paper we will focus on the docking of the enzyme-inhibitor
pairs in the protein docking benchmarks 1 and 2 \cite{benchmark1,benchmark2}. For comparison we also docked the 
antibody-antigen pairs in the benchmark set 1. However, since oligomeric proteins and enzyme-inhibitor 
complexes dominate the training set, the current version of the DARS potential is far from optimal for 
antibody-antigen pairs (and the ''other" types of complexes, not considered here). 

As discussed, to develop the DARS (Decoys as the Reference State) type potential one also needs a 
large set of decoys, i.e., docked structures generated by considering only shape 
complementarity as the target function. We have previously generated 20,000 docked 
conformations for each of the 22 targets of the CAPRI docking experiment \cite{wodak1,wodak2}. We
use these structures as the reference decoy set in the current work. Table I shows the
$18 \times 18$ matrix of interaction energies 
for the resulting DARS potential. The new approach provides clear improvement 
over the Atomic Contact Potential (ACP) \cite{ACE}. As we mentioned, the ACP describes 
relatively well the energetics of a largely hydrophobic interfaces, but 
almost completely ignores the electrostatic interactions. 
In particular,  negative-negative and positive-positive interactions 
($DO^{\delta}$-$DO^{\delta}$ and $RN^{\eta}$-$RN^{\eta}$) are weakly attractive, and 
the $DO^{\delta}$-$RN^{\eta}$ interaction is weakly repulsive \cite{ACE}.  
As shown in Table I, the DARS potential does not suffer 
from these problems, and most parwise interactions have signs as expected. For example, the 
$DO^{\delta}$-$DO^{\delta}$ interaction is strongly repulsive, whereas the $DO^{\delta}$-$RN^{\eta}$ 
interaction is strongly attractive.
The only finding that is somewhat unexpected is the slightly 
attractive $RN^{\eta}$-$RN^{\eta}$ DARS energy, most likely due to the interactions between the hydrophobic 
parts of the arginine side chains. As usual with structure-based potentials, the large value of the 
$SC^{\gamma}$-$SC^{\gamma}$ term, representing cystine-cystine interactions, is an artifact and 
can be ignored. Disulfide bridges in protein-protein 
interfaces are rare, causing this coefficient to be determined from few occurrences. 

We used unbound-unbound (in some cases bound-unbound) 
enzyme-inhibitor and antigen-antibody complex structures of the protein docking benchmark 
1.0 \cite{benchmark1} to test the DARS potential, first for its ability of finding 
''hits", i.e., conformations with less than 10 \AA\ 
$C_{\alpha}$ RMSD from the native, in a large set of docked structures.
Throughout the paper the RMSD is calculated by superimposing the unbound receptor 
(the larger protein) on
the receptor structure in the complex, and calculating the RMSD for the ligand. 
The entry 1TAB was excluded, because it forms a dimer of two complexes, 
in which the carboxy-terminal tail of the inhibitor extends into the interface between
the two trypsin molecules and interacts with both of them simultaneously \cite{1tab}, 
and such multisubunit interactions are not considered in our calculations. 
Table II lists the target complexes. In addition to the Protein Data Bank (PDB) code, we show the 
Coulombic electrostatic 
interaction energy $E_{elec}$ between the two component proteins, calculated 
with the distance-dependent dielectrics $\epsilon = 4r$, as well as $E_{ACP}$, the interaction energy 
through the interface calculated by the Atomic Contact Potential \cite{ACE}. 
As discussed, due to the properties of the ACP \cite{ACE}, a negative $E_{ACP}$ indicates a largely 
hydrophobic interface. 
Thus, Table II shows that in most antibody-antigen complexes the interface is mostly polar. 
The value of $E_{ACP}$ is more variable in enzyme-inhibitor complexes,  but most of these 
have fairly hydrophobic interfaces. 

The unbound proteins for each target listed in Table II are given 
in the original benchmark paper \cite{benchmark1}. 
To test the discriminatory power of structure-based potentials,
we have used these unbound proteins and
the DOT docking program \cite{mandell} with 
a geometrical scoring function to generate 20,000 conformations for each target. The column labeled 20K in 
Table II shows the number of hits among these 20,000 structures. 
For evaluating various scoring functions we used them to rank the 20,000 conformations, 
selected the top 2,000, and determined the number of hits. The better the scoring function, 
the closer we should get to the maximum number of ``hits" among the 20,000 structures or ``decoys", 
shown as 20K. Table II shows first the discrimination results for a ``mixed" strategy we have used 
for many years  \cite{Gatchell,ClusPro}. The strategy involves calculating the electrostatic 
interaction energy 
$E_{elec}$ and the ACP energy $E_{ACP}$ for all the 20,000 structures, and 
retaining 500 with the best (lowest) $E_{ACP}$ values, and an additional 
1500 with the best $E_{elec}$ values. The motivation of this strategy is to have
acceptable discrimination 
for complexes that are stabilized by strong hydrophobic interactions, but also for those that are
not. We keep more structures with favorable electrostatics, since $E_{elec}$ is much more 
sensitive to small perturbations in the coordinates than  $E_{ACP}$. For comparison, Table II
also lists the number of hits within the 2,000 structures with the lowest ACP values, and clearly shows that 
the atomic contact potential is a good discriminator for complexes with a hydrophobic interface
(i.e., when $E_{ACP}$ is negative). However, most of the hits may be lost if $E_{ACP}$ is 
positive, indicating a more polar interface. The ``mixed" strategy improves results for many 
 such complexes, but reduces the number of hits for complexes with a predominantly hydrophobic 
interface. The next column in Table II provides the number of hits among the 2000 conformations 
with the best (lowest) values of the new DARS potential. It appears that 
DARS performs somewhat worse than ACP for complexes with a hydrophobic interface, but it also 
finds hits for the electrostatically driven complexes, achieving a discrimination 
performance which is close to
the one provided by the ``mixed" strategy, although no explicit electrostatic interactions 
were taken into account. This is an important advantage, because a contact potential 
is much less sensitive to small perturbations in the coordinates than the electrostatic energy,  
and this will contribute to improving the docking results. Finally, the last column of Table II (DARS+ACP) 
demonstrates that the results can be further improved by a simple combination of the DARS 
and ACP potentials. Although DARS+ACP does not necessarily produce the best results among the 
considered strategies, 
it does not fail badly for any of the complexes, demonstrating a balanced performance. 

For the docking we calculate the eigenvalues and 
eigenvectors of the matrix of pairwise interactions in the DARS+ACP potential, 
and restrict consideration to the eigenvectors 
corresponding to the few eigenvalues with the largest magnitude. Table III shows the top 4 eigenvalues and 
the corresponding eigenvectors of the combined DARS+ACP potential. 
If we ignore the last column ($CS^{\gamma})$, representing cystine-cystine interactions and based on poor statistics, 
the largest elements in the first eigenvector are for $LC^{\delta}$ and $FC^{\zeta}$, 
both groups representing hydrophobic side chain atoms. 
Thus, the first and largest eigenvalue represents favorable (i.e., negative) hydrophobic interactions. Notice that 
in the second eigenvector the hydrophobic components are small, i.e., the favorable contribution due to
the first (negative) eigenvalue is not 
affected. The second eigenvector shows that Lys side chains (i.e., groups $KN^{\zeta}$ and $KC^{\delta}$) 
are generally not favorable in the interface.  The same vector indicates repulsive same-sign electrostatic 
(i.e., $DO^{\delta}$-$DO^{\delta}$  and $RN^{\eta}$-$RN^{\eta}$) interactions. 
Since the eigenvalues $\lambda_3$ and up are substantially 
smaller in magnitude than $\lambda_1$ and $\lambda_2$, in this paper we restrict consideration to 
the first two eigenvectors. 

{\bf Docking Results} \\
To test the new FFT program we first docked the enzyme-inhibitor pairs from the 
protein docking benchmarks 1.0 and 2.0 
\cite{benchmark1,benchmark2}.  This test set, shown in Table IV, excludes the complexes 
1D6R and 1EWY for the same
reason as 1TAB was excluded, i.e., the complexes are oligomeric, and the intersubunit interactions 
affect the results. Although Table IV lists the PDB codes of the complexes, in the docking calculations
we used the unbound-unbound or the unbound-bound structures of the component proteins as available in 
the benchmark sets \cite{benchmark1,benchmark2}. Notice that for some complexes (e.g., for 1ACB) the two sets 
provide different unbound structures. In such cases the structures given in benchmark set 1 \cite{benchmark1} 
were used. In each calculation the center of mass of 
these component proteins were moved to 
the center of the coordinate system, and were randomly rotated in order to avoid that
the correct docked conformations occur at a grid position. We checked  on several 
complexes that such random perturbations in grid placement can change the results 
by more than 10\%.  

Table IV shows the percentage of the
hits (structures with less than 10 \AA\ $C\alpha$ RMSD from the native) in the top 
1000 and in the top 2000 docked conformations generated by our program PIPER, as well as by 
one of the best FFT-based docking programs ZDOCK 
\cite{ZDOCK}. According to these results, PIPER is a major improvement relative to ZDOCK,
although the latter also works extremely well for enzyme-inhibitor complexes. Generating 2000 structures 
for each of the 33 complexes, 
which is the default for the ZDOCK server, PIPER performs much better (defined as producing at least 50\% more hits) 
than ZDOCK in 14 cases, it is better in 11 cases, worse in 5, and the results are
essentially the same for 3 complexes.
Thus, the results improve in 75\% of the tests, and get worse in  15\%.
Considering only the top 1000 conformations, the improvements are even more pronounced for a 
number of complexes. The results become 
much better in 19 cases, better in 4, worse in 6, and remain essentially unchanged in 4. 
It is particularly advantageous
to have a larger number of hits in the top 1000 structures, because retaining 1000 rather than 2000 structures 
substantially facilitates finding the near-native conformations among them \cite{Gatchell,ClusPro}. 
The improvement is substantial for a number of complexes: for example, the top 1000 structures generated by 
ZDOCK for the complex 1ACB contains 89 hits, whereas the number of hits in 1000 structures 
generated by PIPER is 632. Of course, 
not all cases are this great, but as we stated, the improvement is more than 50 \% over ZDOCK for 19 of the 33 
complexes in Table IV. 
The test set includes one ''difficult" case (1KKL) for which neither method generated any near-native solution. The other
structures for which PIPER did not give good results (e.g., 1BRS and 1DFJ) are nontypical enzyme-inhibitor
complexes in which the association is driven by electrostatics rather than 
hydrophobic shape complementarity. In fact, the Atomic 
Contact Potential (ACP) is generally positive in these complexes (see Table II). It is clear that our 
current DARS potential is far from optimal if the interface is not hydrophobic enough, and  
further development is required for this case. 

Table IV also shows the percentage of hits in the 2000 structures 
generated when using only shape complementarity as the scoring function. Finally, the last column 
labeled `Filter' shows the results of generating 20,000 conformations using the shape complementarity 
part as the scoring function, and then selecting among them the 2000 conformations with the lowest values of 
the complete scoring function, including electrostatics and the DARS potential.  
>From these results it is clear that including the latter energy terms in the  
docking stage yields much better results than docking first for good shape 
complementarity, and then re-ranking and filtering with the additional energy terms. 
In fact, for enzyme-inhibitor complexes even the top 70,000 decoys generated using only 
shape complementarity include fewer
"hits" than the 2000 structures generated by using the complete scoring function in the docking. 
Figure 2 shows, for the enzyme-inhibitor complexes, how the number of hits 
depends on the number of docked structures retained. For comparison, the same curves are also 
shown for ZDOCK and for the use of shape complementarity term as the scoring function.

For the antigen-antibody docking problems we restricted consideration to the complexes in 
benchmark set 1.0. The results shown in Table V and Fig. 3, although comparable to those obtained by 
other docking methods, should be considered preliminary. PIPER yields more hits than ZDOCK in 12 of the 16 test 
problems, but the improvements are much less substantial than the ones we have seen for enzyme-inhibitor
complexes. In fact, we already noted that the current version of 
the DARS potential is far from optimal for antibody-antigen complexes.
This is not surprising, as analyses of protein complexes \cite{jones, LoConte, jackson, chakrabarti}  show that 
the interfaces in enzyme-inhibitor and antibody-antigen complexes substantially differ \cite{glass}. In particular, the latter 
interfaces are generally more polar, more planar, less well packed, and include more water molecules 
than the enzyme-inhibitor interfaces, and these differences result in more challenging docking and free energy evaluation 
problems. The analysis of docking results from three different research groups clearly shows this increased
level of difficulty, even when the the segments that belong to the Complementarity Determining Regions (CDRs) are 
{\it a priori} known \cite{glass}. Therefore,  we are convinced that the results can
be substantially improved by introducing a potential specific to these pairs. 
However, since the number of antigen-antibody complex structures in the PDB is relatively small, 
the development of antigen-antibody potentials will require reducing the number of 
atom types from 18, a topic of our current research. Nevertheless, the
antibody-antigen docking results shown in Table V are good enough to indicate that the use of pairwise potentials 
in docking increases the 
number of hits among the complex structures generated. 

The program PIPER was implemented in C for different
cluster environments. The CPU time required for determining an average complex by 
docking the free component proteins is 40 minutes on a 
30 dual processor cluster with P3 1GHZ nodes, and it is approximately 
2 minutes on 512 nodes of an IBM BlueGene/L. 
The  PIPER program is free to academic users, and will be sent upon request. 

{\bf CONCLUSIONS}

We have extended the well known Fast Fourier Transform (FFT) correlation approach for use with 
pairwise potentials defined among $K$ different atom types. The method involves the 
eigenvalue-eigenvector decomposition of the $K \times K$
interaction energy matrix, which converts the scoring function to the sum of $K$ correlation functions. 
Although correlation functions can be efficiently evaluated by FFT calculations, the computational costs are 
prohibitive if $K$ is large. The main contribution of the paper is the observation that 
one can restrict consideration to a few (2 to 4) dominant eigenvalues and the corresponding eigenvectors 
of the interaction energy matrix without substantially reducing the accuracy of the method, but substantially reducing
the computational costs and rendering the approach computationally feasible.

We use the new FFT method with a novel structure-based potential termed DARS 
(Decoys As the Reference State), extracted from a set of protein-protein complex structures. 
The novelty of the DARS potential is that we generate large sets of docked structures 
using shape complementarity as the scoring function, and use these structures to derive the frequencies 
of atom pair interactions in the reference state. Since  
the decoy structures do not depend on specific atomic interactions, they can be considered  
random complexes. Thus, the probability of interaction between two atoms of types $I$ and $J$, respectively,
can be estimated by determining the frequency of $I-J$ interactions in protein complexes, divided by 
the frequency of $I-J$ interactions in the decoy set. Since our goal is finding 
docked structures with high levels of ``chemical" complementarity among the many compact structures 
generated by the FFT algorithm, it is natural to define the probability 
of interaction between two atoms as the
frequency of the pair in native complexes, divided by the 
frequency of the same pair in the decoys. We have found the best performance using a 
linear combination of the new 
DARS potential and the Atomic Contact Potential (ACP) \cite{ACE}, an atom-level extension 
of the Miyazawa-Jernigan potential. Note that although the results of this paper are based on 
the use of the combined DARS+ACP potential, the 
new FFT method can be used with any pairwise potential as part of the scoring function.   

The method has been tested on docking enzyme-inhibitor and antibody-antigen pairs. 
It was expected that the use of pairwise potentials would  improve 
the results. We have found that the improvement is substantial for enzyme-inhibitor complexes, whose 
energetics is well described by the current version of the 
DARS+ACP potential. Indeed, for 19 of the the 33 enzyme-inhibitor 
pairs considered, the number of ``hits" (near-native structures) 
in the top 1000 docked conformations has been increased by more than 50\% relative 
to ZDOCK, one of the best FFT-based docking programs \cite{ZDOCK}. 
Although the improvements are 
less substantive for antigen-antibody complexes, the results show that, due to the use of 
pairwise potentials, the new program PIPER tends to produce more hits 
than traditional FFT-based methods. Our results clearly show that the improvement is due to the 
use of the pairwise potential directly in the docking calculations. In fact, the 
two-step strategy of 
generating a large number of docked conformations and then ranking them with the 
pairwise potential yields much fewer near-native structures.  We believe that the results for antibody-antigen 
pairs can be further improved by developing a specific potential which is more appropriate for this type 
of complexes. 

Since we need only a few eigenvectors to estimate the pairwise interaction matrix, the computational load is
relatively moderate. This implies that the method will be applicable not only to contact potentials considered 
in the present work but also to distance dependent potentials.
The use of more detailed scoring 
functions is expected to further improve the results. We note that the distance-dependent potentials will be represented 
as sums of pairwise contact potentials with different cutoff radii, and hence they will increase the number of 
required FFT calculations. However, with  computer speeds consistently increasing, the approach will remain 
computationally feasible. Thus, the primary limitation on further improving the method is the accuracy of
the potential functions. This accuracy in part is determined by the availability of 
protein-protein complex structures, which is expected to grow.  

We note that in the past pairwise potentials have been used with great success in the second step of 
docking for finding near-native docked conformations among the thousands of structures generated. As
shown here, it is much more effective to use such pairwise potentials directly in the docking step 
rather than for discrimination. Indeed, 
we have shown that the top 1000 structures from the docking generally include a fair number of near-native 
complex conformations. However, it is still necessary to identify the best models among these 1000 retained.  
It is clear that we can not use the same potential that has been used for docking. More generally, we have recently 
shown that the combination of different potentials can substantially improve docking and discrimination 
results \cite{Murphy}. This fact emphasizes the need for developing higher accuracy potentials that combine molecular 
mechanics with empirical solvation and entropic terms, and are able to discriminate near-native complex 
conformations from the rest of structures generated by the docking. 

{\bf ACKNOWLEDGMENTS}

The development of the PIPER program has been partially supported by SolMap Pharmaceuticals, Inc., 
in collaboration with Mercury Computer 
Systems, Inc. The application of the program to protein-protein docking and the development of 
structure-based potentials have been supported 
by grant GM61867 from the National Institute of Health. For the CPU time used for this paper we are grateful 
to grant MRI DBI-0116574 and to the Boston University Scientific Computing and Visualization Center 
for the opportunity of running the program on the Blue Gene/L supercomputer.

\newpage
\bibliography{fft} 

\newpage
\section{Figure Legends}
{\bf Fig. 1}
(A) Slice of the scalar electrostatic potential field of Ribonuclease A, the receptor in the complex 1DFJ. 
(B) The same slice as in figure (A), but the potential is smoothed by convolution as described in Methods.
(C) Slice of the field of the electrostatic receptor-ligand interaction in the complex of ribonuclease A with ribonuclease 
inhibitor (1DFJ) as function of the translations of the inhibitor along 2 coordinates. The ligand orientation is 
as in the native complex. The large spike indicates the native  position of the ligand. (D) The same as in (C),
but the electrostatic interaction energy  is calculated using the smoothed potential.
\\\\
{\bf{Fig. 2}}
The number of hits (near-native structures with less than 10 \AA\ 
RMSD from the native) as function of the number of docked conformations retained from the FFT
calculations for the enzyme-inhibitor complexes in the protein-protein docking benchmark sets 1 and 2. 
The curves are coded as follows: red crosses -- PIPER; green stars -- ZDOCK, version 2.3; blue -- predictions 
using shape complementarity as the scoring function. 
\\\\
{\bf{Fig. 3}}\\
The number of hits (near-native structures with less than 10 \AA\ 
RMSD from the native) as function of the number of docked conformations retained from the FFT
calculations for the antibody-antigen complexes in the protein-protein docking benchmark 1. 
Residues not in the Complementarity Determining Regions (CDRs) were masked by removing the 
attractive shape complementarity term. Color codes are as in Fig. 2. 
\\\\
\newpage
\begin{singlespace}
\begin{minipage}[t]{\linewidth}
\begin{center}
\begin{tabular}{cccccc}
\resizebox{20mm}{!}{\includegraphics{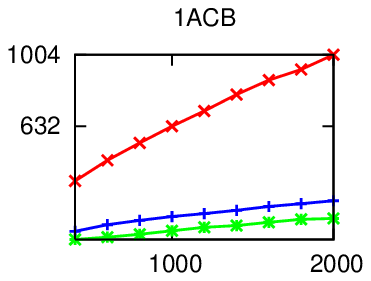}}&
\resizebox{20mm}{!}{\includegraphics{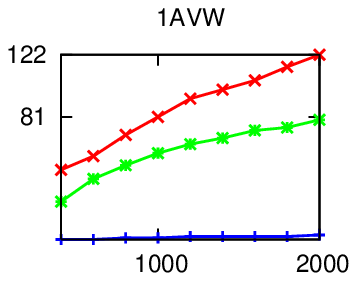}}&
\resizebox{20mm}{!}{\includegraphics{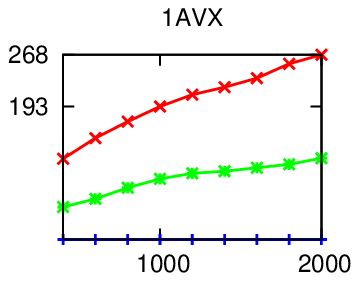}}&
\resizebox{20mm}{!}{\includegraphics{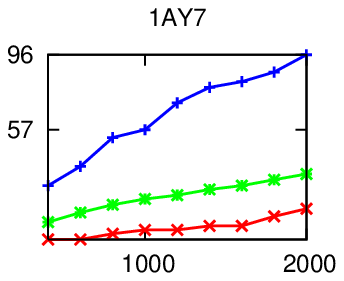}}&
\resizebox{20mm}{!}{\includegraphics{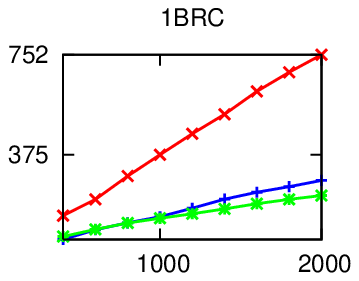}}&
\resizebox{20mm}{!}{\includegraphics{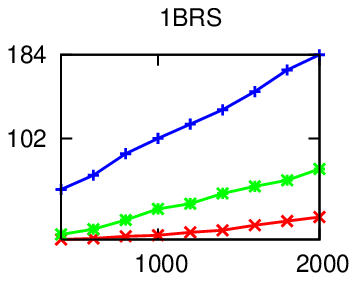}}\\
\resizebox{20mm}{!}{\includegraphics{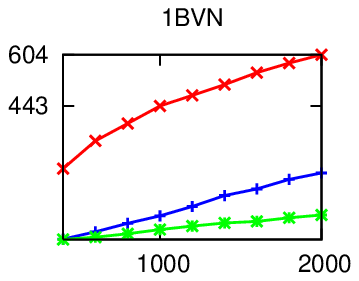}}&
\resizebox{20mm}{!}{\includegraphics{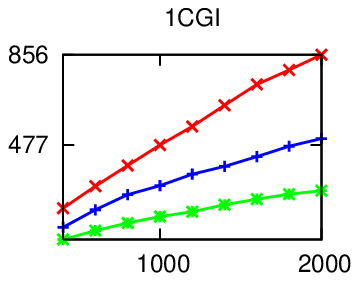}}&
\resizebox{20mm}{!}{\includegraphics{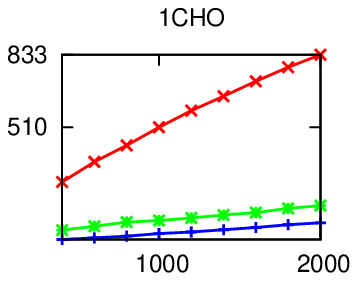}}&
\resizebox{20mm}{!}{\includegraphics{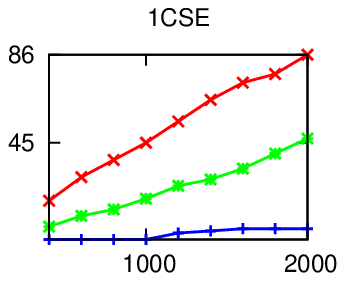}}&
\resizebox{20mm}{!}{\includegraphics{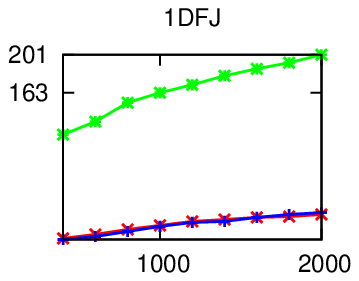}}&
\resizebox{20mm}{!}{\includegraphics{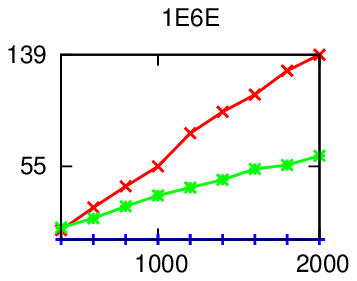}}\\
\resizebox{20mm}{!}{\includegraphics{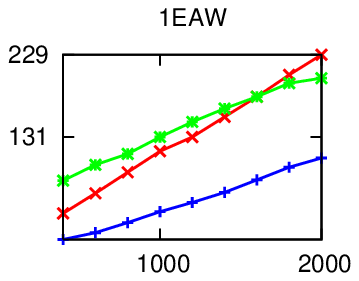}}&
\resizebox{20mm}{!}{\includegraphics{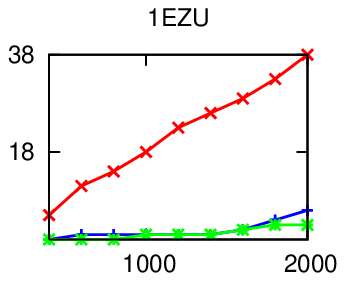}}&
\resizebox{20mm}{!}{\includegraphics{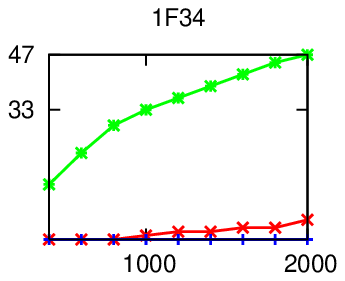}}&
\resizebox{20mm}{!}{\includegraphics{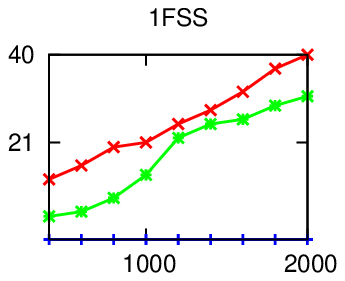}}&
\resizebox{20mm}{!}{\includegraphics{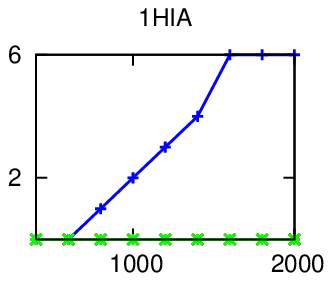}}&
\resizebox{20mm}{!}{\includegraphics{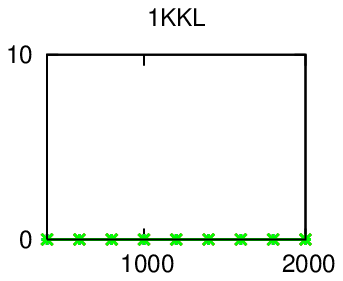}}\\
\resizebox{20mm}{!}{\includegraphics{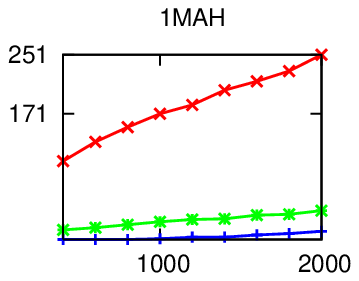}}&
\resizebox{20mm}{!}{\includegraphics{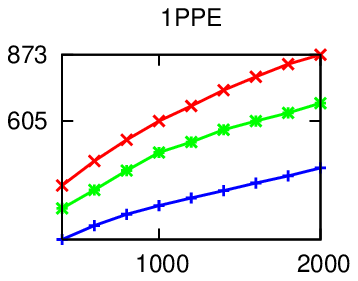}}&
\resizebox{20mm}{!}{\includegraphics{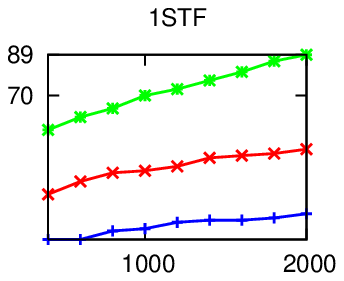}}&
\resizebox{20mm}{!}{\includegraphics{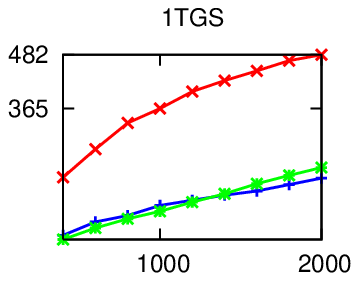}}&
\resizebox{20mm}{!}{\includegraphics{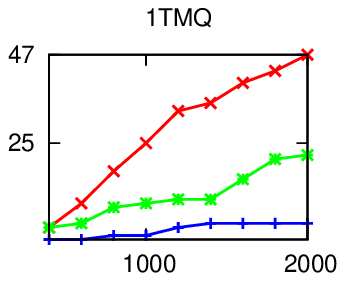}}&
\resizebox{20mm}{!}{\includegraphics{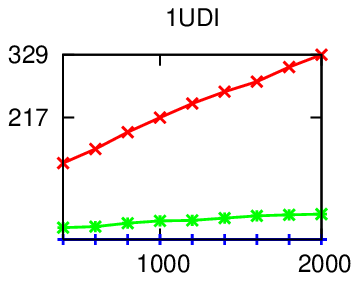}}\\
\resizebox{20mm}{!}{\includegraphics{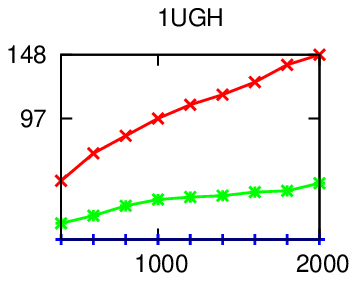}}&
\resizebox{20mm}{!}{\includegraphics{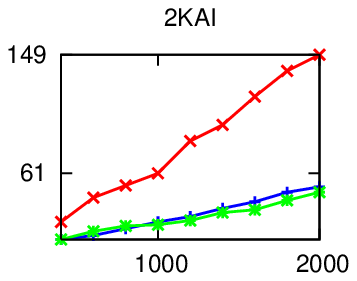}}&
\resizebox{20mm}{!}{\includegraphics{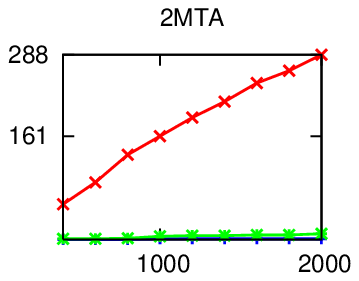}}&
\resizebox{20mm}{!}{\includegraphics{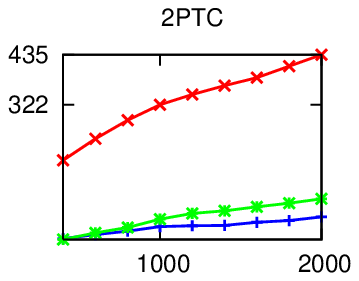}}&
\resizebox{20mm}{!}{\includegraphics{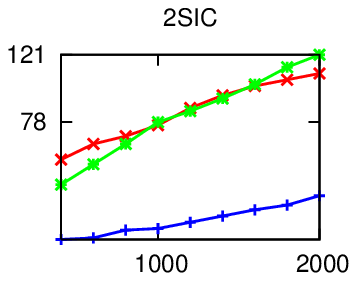}}&
\resizebox{20mm}{!}{\includegraphics{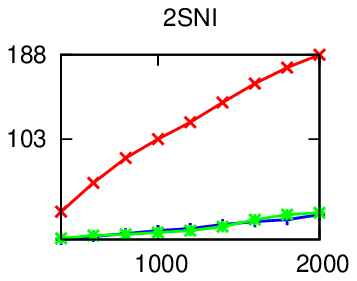}}\\
\resizebox{20mm}{!}{\includegraphics{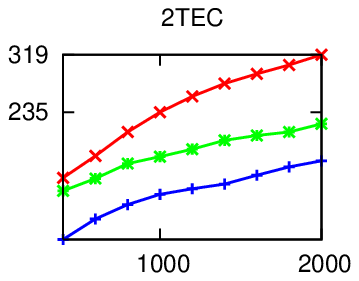}}&
\resizebox{20mm}{!}{\includegraphics{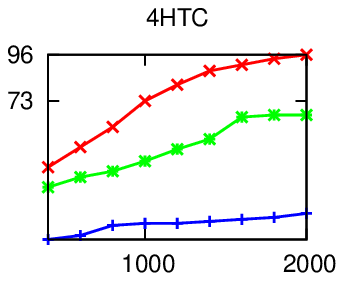}}&
\resizebox{20mm}{!}{\includegraphics{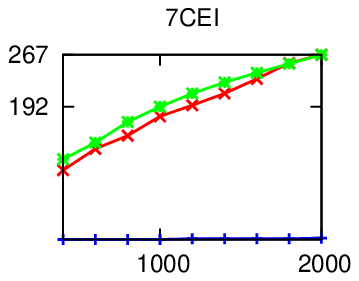}}&
\hspace{1pt}&\hspace{1pt}&\hspace{1pt}\\
\end{tabular}
\label{fig:ei}
\figcaption{}
 \end{center}
\end{minipage}
\newpage
\begin{minipage}[t]{\linewidth}
\begin{center}
 \begin{tabular}{cccccc}
\resizebox{20mm}{!}{\includegraphics{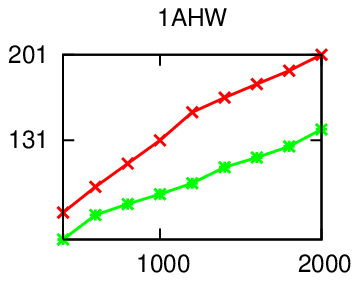}}&
\resizebox{20mm}{!}{\includegraphics{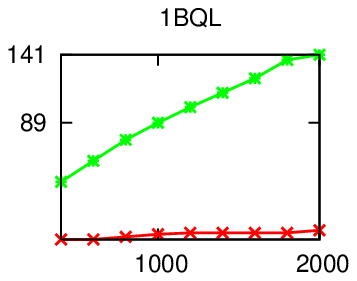}}&
\resizebox{20mm}{!}{\includegraphics{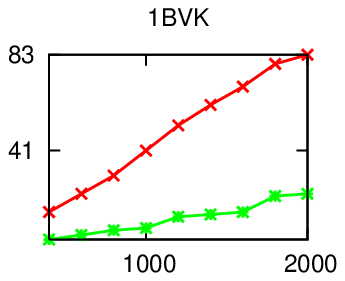}}&
\resizebox{20mm}{!}{\includegraphics{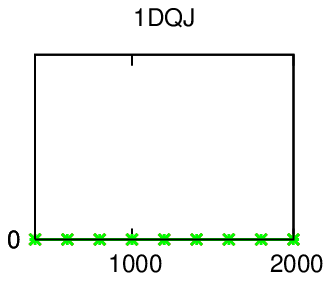}}&
\resizebox{20mm}{!}{\includegraphics{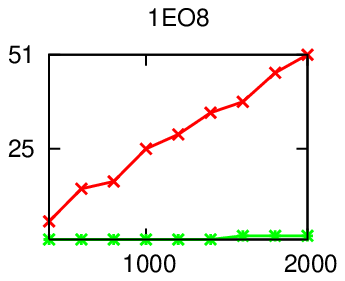}}&
\resizebox{20mm}{!}{\includegraphics{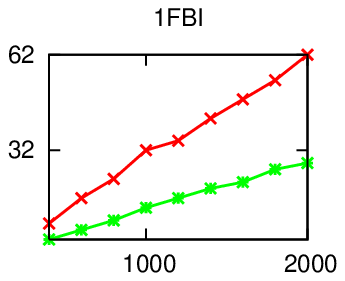}}\\
\resizebox{20mm}{!}{\includegraphics{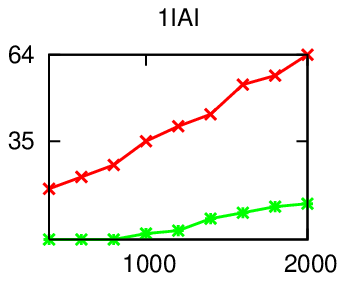}}&
\resizebox{20mm}{!}{\includegraphics{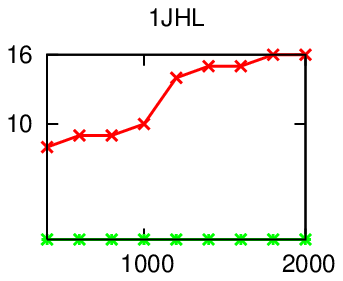}}&
\resizebox{20mm}{!}{\includegraphics{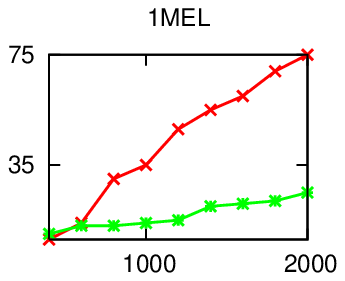}}&
\resizebox{20mm}{!}{\includegraphics{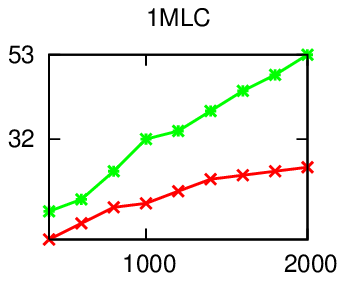}}&
\resizebox{20mm}{!}{\includegraphics{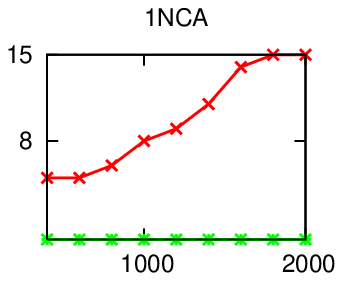}}&
\resizebox{20mm}{!}{\includegraphics{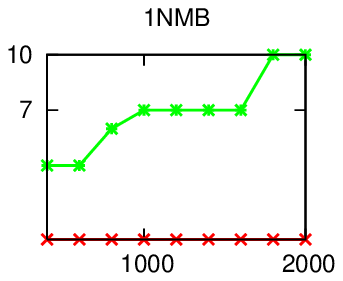}}\\
\resizebox{20mm}{!}{\includegraphics{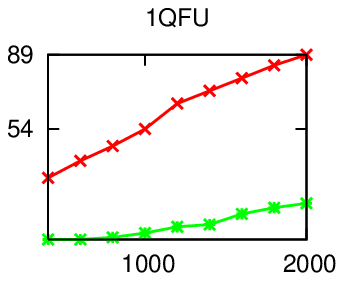}}&
\resizebox{20mm}{!}{\includegraphics{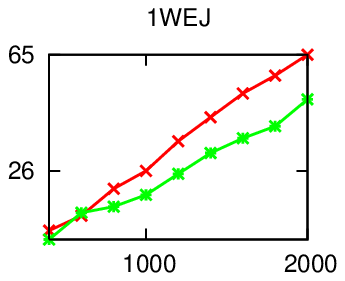}}&
\resizebox{20mm}{!}{\includegraphics{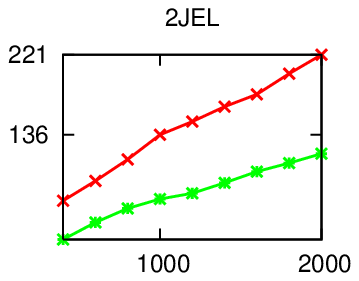}}&
\resizebox{20mm}{!}{\includegraphics{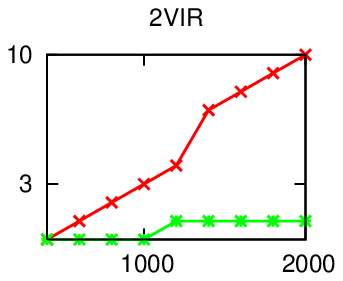}}&
\hspace{1pt}&\hspace{1pt}\\
\end{tabular}
\label{fig:aa}
\figcaption{}
\end{center}
\end{minipage}
\newpage
\begin{landscape}
\begin{center}
\begin{tabular}{ccccccccccccccccccc}
\multicolumn{19}{c}{\bf TABLE I. Contact Energies of the DARS Potential} \\
\hline
&$N$& $C^{\alpha}$ & $C$ & $O$& $GC^{\alpha}$ & $C^{\beta}$ & $KN^{\zeta}$ & $KC^{\delta}$ & $DO^{\delta}$ & 
$RN^{\eta}$ & $NN^{\delta}$ & $RN^{\varepsilon}$ & $SO^{\gamma}$ & $HN^{\varepsilon}$ & $YC^{\zeta}$ & $FC^{\zeta}$ 
& $LC^{\delta}$ & $CS^{\gamma}$ \\
\hline
$N$ & 0.07& 0.03& 0.08& 0.17& 0.23& 0.01& 0.98& 0.93& 0.72& 0.02& 0.38& 0.10& 0.42& -0.05& 0.04& -0.22& -0.60& -1.52\\
$C^{\alpha}$ & 0.03& -0.04& 0.05& 0.18& 0.20& -0.11& 0.83& 0.78& 0.67& -0.08& 0.30& -0.04& 0.34& -0.18& -0.11& -0.38& -0.79& -1.54\\
$C$ & 0.08& 0.05& 0.05& 0.12& 0.29& 0.01& 0.92& 0.76& 0.71& -0.03& 0.35& -0.01& 0.41& -0.12& 0.00& -0.26& -0.62& -1.54\\
$O$ & 0.17& 0.18& 0.12& 0.12& 0.40& 0.06& 0.85& 0.71& 0.73& -0.01& 0.37& -0.12& 0.45& -0.14& 0.04& -0.25& -0.60& -1.40\\
$GC^{\alpha}$ & 0.23& 0.20& 0.29& 0.40& 0.17& 0.13& 0.80& 0.68& 0.77& -0.00& 0.59& -0.06& 0.66& -0.10& 0.18& -0.11& -0.28& -0.88\\
$C^{\beta}$ & 0.01& -0.11& 0.01& 0.06& 0.13& -0.25& 0.75& 0.69& 0.49& -0.12& 0.22& -0.13& 0.26& -0.26& -0.26& -0.54& -0.95& -1.52\\
$KN^{\zeta}$ & 0.98& 0.83& 0.92& 0.85& 0.80& 0.75& 1.20& 1.34& 0.12& 0.66& 0.81& 1.00& 0.90& 0.85& 0.61& 0.71& 0.70& 0.19\\
$KC^{\delta}$ & 0.93& 0.78& 0.76& 0.71& 0.68& 0.69& 1.34& 1.37& 0.19& 0.79& 0.82& 1.06& 0.91& 0.87& 0.42& 0.48& 0.28& 0.04\\
$DO^{\delta}$ & 0.72& 0.67& 0.71& 0.73& 0.77& 0.49& 0.12& 0.19& 0.70& -0.55& 0.54& -0.46& 0.45& -0.15& 0.19& 0.31& 0.21& 0.06\\
$RN^{\eta}$ & 0.02& -0.08& -0.03& -0.01& -0.00& -0.12& 0.66& 0.79& -0.55& -0.36& 0.23& -0.46& 0.20& -0.26& -0.21& -0.28& -0.30& -0.27\\
$NN^{\delta}$ & 0.38& 0.30& 0.35& 0.37& 0.59& 0.22& 0.81& 0.82& 0.54& 0.23& 0.13& 0.24& 0.53& 0.23& 0.18& 0.09& -0.10& -0.07\\
$RN^{\varepsilon}$ & 0.10& -0.04& -0.01& -0.12& -0.06& -0.13& 1.00& 1.06& -0.46& -0.46& 0.24& -0.50& 0.19& -0.34& -0.46& -0.47& -0.56& -0.68\\
$SO^{\gamma}$ & 0.42& 0.34& 0.41& 0.45& 0.66& 0.26& 0.90& 0.91& 0.45& 0.20& 0.53& 0.19& 0.45& -0.00& 0.22& 0.09& -0.20& -0.79\\
$HN^{\varepsilon}$ & -0.05& -0.18& -0.12& -0.14& -0.10& -0.26& 0.85& 0.87& -0.15& -0.26& 0.23& -0.34& -0.00& -0.93& -0.46& -0.40& -0.76& -1.67\\
$YC^{\zeta}$ & 0.04& -0.11& 0.00& 0.04& 0.18& -0.26& 0.61& 0.42& 0.19& -0.21& 0.18& -0.46& 0.22& -0.46& -0.31& -0.48& -0.80& -1.39\\
$FC^{\zeta}$ & -0.22& -0.38& -0.26& -0.25& -0.11& -0.54& 0.71& 0.48& 0.31& -0.28& 0.09& -0.47& 0.09& -0.40& -0.48& -0.96& -1.31& -1.46\\
$LC^{\delta}$ & -0.60& -0.79& -0.62& -0.60& -0.28& -0.95& 0.70& 0.28& 0.21& -0.30& -0.10& -0.56& -0.20& -0.76& -0.80& -1.31& -1.98& -2.00\\
$CS^{\gamma}$ & -1.52& -1.54& -1.54& -1.40& -0.88& -1.52& 0.19& 0.04& 0.06& -0.27& -0.07& -0.68& -0.79& -1.67& -1.39& -1.46& -2.00& -5.67\\
\hline
\end{tabular}
\end{center}
\end{landscape}
\newpage
\centering
{\bf TABLE II. Number of Hits Retained by Various Scoring Functions}
\begin{tabular}{lcrrrrrrr}
\hline
\multirow{2}{*}{Complex} & \multirow{2}{*}{Type$^a$} & \multicolumn{2}{c}{Binding
  free energy$^b$}
&\multirow{2}{*}{20K$^c$}&\multicolumn{4}{c}{2K$^d$}\\ 
\cline{3-4}\cline{6-9}
  & &$E_{elec}$ &$E_{ACP}$ & & Mixed$^e$ & ACP$^f$
  & DARS$^g$ & DARS+ACP$^h$\\
\hline
1ACB & e-i&  -12.23 &  -16.08 &  218 &  99 &  168 & 135 & 132 \\
1AVW & e-i&  -24.39 &  -0.71 &  44 &  2 &  3 & 4 & 4 \\
1BRC & e-i&  -15.08 &  -7.61 &  175 &  39 &  80 & 115 & 108 \\
1BRS & e-i&  -41.17 &  11.59 &  113 &  25 &  5 & 25 & 27 \\
1CGI & e-i&  -14.46 &  -18.29 &  161 &  36 &  123 & 27 & 28 \\
1CHO & e-i&  -14.10 &  -12.28 &  183 &  66 &  101 & 91 & 90 \\
1CSE & e-i&  -18.40 &  -8.56 &  248 &  107 &  201 & 64 & 80 \\
1DFJ & e-i&  -63.93 &  18.75 &  31 &  30 &  0 & 30 & 30 \\
1FSS & e-i&  -35.35 &  -0.13 &  15 &  6 &  7 & 5 & 5 \\
1MAH & e-i&  -30.01 &  -4.47 &  12 &  8 &  10 & 8 & 8 \\
1PPE & e-i&  -19.62 &  -7.22 &  354 &  124 &  145 & 174 & 168 \\
1STF & e-i&  -6.33 &  -10.25 &  71 &  22 &  36 & 17 & 20 \\
1TGS & e-i&  -23.48 &  -8.37 &  186 &  57 &  153 & 44 & 55 \\
1UDI & e-i&  -35.86 &  -0.53 &  102 &  41 &  78 & 44 & 44 \\
1UGH & e-i&  -37.84 &  0.72 &  115 &  34 &  54 & 46 & 44 \\
2PTC & e-i&  -22.87 &  -3.76 &  153 &  57 &  132 & 33 & 46 \\
2SIC & e-i&  -11.74 &  -14.01 &  89 &  66 &  84 & 52 & 61 \\
2SNI & e-i&  -12.18 &  -12.61 &  127 &  56 &  92 & 55 & 58 \\
2TEC & e-i&  -11.67 &  -11.17 &  270 &  131 &  166 & 70 & 75 \\
4HTC & e-i&  -53.05 &  4.07 &  118 &  10 &  7 & 47 & 38 \\
1AHW & a-a&  -44.05 &  18.89 &  186 &  93 &  0 & 93 & 93 \\
1BQL & a-a&  -34.52 &  7.58 &  280 &  43 &  49 & 39 & 39 \\
1BVK & a-a&  -10.98 &  9.20 &  234 &  15 &  33 & 22 & 25 \\
1DQJ & a-a&  -20.51 &  13.63 &  35 &  1 &  0 & 1 & 1 \\
1E08 & a-a&  -21.39 &  -0.40 &  90 &  22 &  10 & 55 & 53 \\
1FBI & a-a&  -38.82 &  10.80 &  45 &  7 &  3 & 7 & 7 \\
1IAI & a-a&  -5.90 &  -2.22 &  139 &  30 &  82 & 56 & 57 \\
1JHL & a-a&  -19.67 &  10.05 &  51 &  10 &  0 & 10 & 10 \\
1MLC & a-a&  -21.44 &  -3.84 &  165 &  40 &  78 & 32 & 48 \\
1NCA & a-a&  -30.96 &  8.41 &  72 &  13 &  0 & 13 & 13 \\
1NMB & a-a&  -20.67 &  1.38 &  19 &  8 &  3 & 8 & 8 \\
1QFU & a-a&  -20.75 &  -0.74 &  126 &  16 &  45 & 48 & 42 \\
1WEJ & a-a&  -37.62 &  15.92 &  20 &  20 &  0 & 20 & 20 \\
2JEL & a-a&  -14.50 &  8.42 &  74 &  7 &  2 & 8 & 7 \\
2VIR & a-a&  -16.99 &  1.10 &  118 &  15 &  40 & 12 & 12 \\
\hline
\multicolumn{9}{l}{$^a$e-i: enzyme-inhibitor; a-a: antibody-antigen}\\
\multicolumn{9}{l}{$^b$Based on bound complex conformation}\\
\multicolumn{9}{l}{$^c$20,000 structures with the best shape complementarity, generated by DOT}\\
\multicolumn{9}{l}{$^d$2,000 structures selected from the 20,000 generated by DOT}\\
\multicolumn{9}{l}{$^e$500 structures with the best $E_{ACP}$ and 1500 structures with the best $E_{elec}$}\\
\multicolumn{9}{l}{$^f$2,000 structures with the best $E_{ACP}$}\\  
\multicolumn{9}{l}{$^g$2,000 structures with the best values of the DARS potential}\\  
\multicolumn{9}{l}{$^h$2,000 structures with the best values of the DARS+ACP potential}\\  
\end{tabular}
\newpage
\begin{landscape}
\begin{center}
\begin{tabular}{ccccccccccccccccccc}
\multicolumn{19}{c}{\bf TABLE III. Top eigenvalues and eigenvectors for the DARS+ACP Potential}\\
\hline
$\lambda^a$ & $N$& $C^{\alpha}$ & $C$ & $O$& $GC^{\alpha}$ & $C^{\beta}$ & $KN^{\zeta}$ & $KC^{\delta}$ & $DO^{\delta}$ & 
$RN^{\eta}$ & $NN^{\delta}$ & $RN^{\varepsilon}$ & $SO^{\gamma}$ & $HN^{\varepsilon}$ & $YC^{\zeta}$ & $FC^{\zeta}$ 
& $LC^{\delta}$ & $CS^{\gamma}$ \\
\hline
-8.9& -0.17& -0.19& -0.18& -0.15& -0.08& -0.20& 0.16& 0.11& 0.08& -0.05& 0.02& -0.11& -0.05& -0.21& -0.18& -0.25& -0.37& -0.71\\
6.6 & -0.21& -0.17& -0.19& -0.22& -0.23& -0.15& -0.48& -0.47& -0.21& -0.15&
-0.26& -0.18& -0.28& -0.13& -0.10& -0.07& 0.02& 0.18\\
1.8& -0.08& -0.13& -0.14& -0.22& -0.21& -0.09& 0.26& 0.32& -0.67& 0.29& -0.05& 0.31& -0.09& 0.12& -0.12& -0.03& 0.07& 0.13\\
-1.7& -0.09& -0.01& -0.07& -0.05& -0.02& 0.07& -0.28& -0.22& 0.13& 0.36& 0.02& 0.46& -0.07& 0.16& 0.16& 0.28& 0.33& -0.50\\
\hline
\multicolumn{19}{l}{$^a$Eigenvales with the largest magnitude and the corresponding eigenvectors}\\
\end{tabular}
\end{center}
\end{landscape}
\newpage
\begin{center}
\begin{tabular}{crrrrrr}
\multicolumn{7}{c}{\bf TABLE IV. Percentage of Hits Among Conformations Generated}\\
\multicolumn{7}{c}{\bf by PIPER and ZDOCK for Enzyme-Inhibitor Complexes}\\
\hline
\multirow{2}{*}{Complex} & \multicolumn{2}{c}{1000 predictions}&\multicolumn{4}{c}{2000 predictions}\\
\cline{2-7}
      & PIPER & ZDOCK & \qquad PIPER  & ZDOCK & Shape$^a$ & Filter$^b$ \\
\hline    
1ACB & 63.2 &  8.9 &  50.2 & 7.3 & 16.3 & 8.4\\
1AVW & 8.1 &  5.7 &  6.1 & 4.0 & 0.1 & 0.5\\
1AVX & 19.3 &  8.8 &  13.4 & 5.9 & 0.0 & 0.7\\
1AY7 & 0.5 &  2.1 &  0.8 & 1.7 & 5.7 & 2.7\\
1BRC & 37.5 &  13.5 &  37.6 & 11.1 & 14.1 & 9.8\\
1BRS & 0.7 &  3.3 &  1.3 & 3.6 & 10.2 & 5.7\\
1BVN & 44.3 &  5.4 &  30.2 & 5.0 & 9.7 & 7.5\\
1CGI & 47.7 &  17.6 &  42.8 & 14.2 & 30.6 & 11.8\\
1CHO & 51.0 &  9.5 &  41.7 & 7.9 & 3.7 & 3.9\\
1CSE & 4.5 &  1.9 &  4.3 & 2.4 & 0.0 & 1.4\\
1DFJ & 3.0 &  16.3 &  2.1 & 10.1 & 2.9 & 1.4\\
1E6E & 5.5 &  3.3 &  7.0 & 3.1 & 0.0 & 0.0\\
1EAW & 11.4 &  13.1 &  11.5 & 10.0 & 4.2 & 6.4\\
1EZU & 1.8 &  0.1 &  1.9 & 0.1 & 0.1 & 0.2\\
1F34 & 0.1 &  3.3 &  0.3 & 2.3 & 0.0 & 0.0\\
1FSS & 2.1 &  1.4 &  2.0 & 1.5 & 0.0 & 0.1\\
1HIA & 0.0 &  0.0 &  0.0 & 0.0 & 0.2 & 0.5\\
1KKL & 0.0 &  0.0 &  0.0 & 0.0 & 0.0 & 0.0\\
1MAH & 17.1 &  2.5 &  12.6 & 1.9 & 0.2 & 0.9\\
1PPE & 60.5 &  47.8 &  43.7 & 33.9 & 26.3 & 11.9\\
1STF & 3.5 &  7.0 &  2.3 & 4.4 & 0.8 & 0.5\\
1TGS & 36.5 &  14.2 &  24.1 & 11.8 & 15.5 & 5.5\\
1TMQ & 2.5 &  1.0 &  2.4 & 1.1 & 0.2 & 0.2\\
1UDI & 21.7 &  3.3 &  16.5 & 2.2 & 0.0 & 0.4\\
1UGH & 9.7 &  3.2 &  7.4 & 2.2 & 0.0 & 0.1\\
2KAI & 6.1 &  2.3 &  7.5 & 2.4 & 2.5 & 2.5\\
2MTA & 16.1 &  0.5 &  14.4 & 0.4 & 0.1 & 0.0\\
2PTC & 32.2 &  6.3 &  21.8 & 5.5 & 4.6 & 2.5\\
2SIC & 7.6 &  7.8 &  5.5 & 5.9 & 1.0 & 1.1\\
2SNI & 10.3 &  0.9 &  9.4 & 1.5 & 1.1 & 1.5\\
2TEC & 23.5 &  17.0 &  16.0 & 10.8 & 11.5 & 5.2\\
4HTC & 7.3 &  4.3 &  4.8 & 3.3 & 1.2 & 0.7\\
7CEI & 17.8 &  19.2 &  13.4 & 13.3 & 0.0 & 0.6\\
\hline
\multicolumn{7}{l}{$^a$ Top 2000 structures generated using shape complementarity}\\
\multicolumn{7}{l}{$^b$ The 2000 best scoring structures selected from the 20,000 with the best}\\ 
\multicolumn{7}{l}{\quad shape complemetarity}\\
\end{tabular}
\end{center}
\newpage
\begin{center}
\begin{tabular}{crrrr}
\multicolumn{5}{c}{\bf TABLE V. Percentage of Hits Generated}\\
\multicolumn{5}{c}{\bf for Antibody-Antigen Complexes}\\
\hline
\multirow{2}{*}{Compex} & \multicolumn{2}{c}{1000 predictions}&\multicolumn{2}{c}{2000 predictions}\\
\cline{2-5}
      & PIPER & ZDOCK & \qquad PIPER  & ZDOCK\\
\hline    
1AHW & 13.1 &  8.7 &  10.0 & 7.0\\
1BQL & 0.4 &  8.9 &  0.4 & 7.0\\
1BVK & 4.1 &  0.7 &  4.2 & 1.1\\
1DQJ & 0.0 &  0.0 &  0.0 & 0.0\\
1EO8 & 2.5 &  0.0 &  2.5 & 0.1\\
1FBI & 3.2 &  1.4 &  3.1 & 1.4\\
1IAI & 3.5 &  0.4 &  3.2 & 0.7\\
1JHL & 1.0 &  0.0 &  0.8 & 0.0\\
1MEL & 3.5 &  1.4 &  3.8 & 1.3\\
1MLC & 1.6 &  3.2 &  1.3 & 2.7\\
1NCA & 0.8 &  0.0 &  0.8 & 0.0\\
1NMB & 0.0 &  0.7 &  0.0 & 0.5\\
1QFU & 5.4 &  0.5 &  4.5 & 1.0\\
1WEJ & 2.6 &  1.8 &  3.3 & 2.5\\
2JEL & 13.6 &  6.8 &  11.1 & 5.8\\
2VIR & 0.3 &  0.0 &  0.5 & 0.1\\
\hline
\end{tabular}
\end{center}
\end{singlespace}
\end{document}